\documentclass[11pt]{article}

\usepackage{amsmath,amssymb,amsthm}
\usepackage{geometry}
\usepackage{hyperref}
\usepackage{cite}
\usepackage{bbm}
\usepackage{mathtools}

\title{\large\bf Where in the Island is the Information?}

\author{Tom Banks\\
NHETC and Department of Physics \\
Rutgers University, Piscataway, NJ 08854-8019\\
E-mail: \href{mailto:tibanks@ucsc.edu}{tibanks@ucsc.edu}
\\
\\
}

\date{}

\begin{document}
\maketitle

\begin{abstract}
The Island Formula for Quantum Extremal Surfaces of large AdS black holes, which are made to evaporate by coupling the CFT of the AdS/CFT correspondence to an external system, demonstrates that the space time picture can reproduce the quantum mechanically required Page curve of the CFT/bath system if a region in the interior of the black hole has become entangled with the external radiation by the Page time. In an effective quantum field theory (EFT) description in ``nice slice" coordinates, this appears to imply paradoxical entanglement between degrees of freedom that have been causally disconnected since long before the Page time. We will give two arguments that all of the quantum information in the island is in fact concentrated in a tiny space-like portion of the island very close to the black hole horizon at the Page time.  The first applies the covariant entropy bound\cite{fsb} (CEB) to the causal diamonds of geodesics originating on the space-like region that becomes the island, but on a time slice just before the Page time.  All points on the island except those in a stretched horizon region very close to the horizon have entropy bounds that are tiny fractions of the black hole entropy.  The second argument uses standard quantum statistical mechanics combined with the tensor network renormalization group regularization of CFT to argue that the wave function of a typical pure state in the microcanonical evaporating black hole always has most of its probability concentrated in a small range of wavelengths corresponding to a small range of radii in the dual tensor network/AdS slice.  Thus the wave function that is entangled with the bath after the Page time lives close to the horizon of the adiabatic black hole, within a stretched horizon of width $\sim R_{AdS} (L_P / R_{AdS})^{(d-2)/d}$.   We relate these observations to claims that a localized object hitting the singularity is dual to equilibration of that object's degrees of freedom, with those of the horizon.  The implications for non-isometric encodings of quantum information are pointed out.   
\end{abstract}

\tableofcontents

\section{Introduction}

Maldacena's discovery\cite{aharonyetal} that a large class of large $N$ Super Conformal Field Theories were dual to models of quantum gravity in space-times of the form $AdS_d \times K$ with $K$ a compact manifold and the scale of both manifolds much larger than any microscopic scale like the Planck or string lengths gave us a tool to study properties of quantum gravity using the well understood tools of QFT.  In particular, black holes with radius large compared to $R_{AdS}$ are simply thermal states of the CFT.  These are stable, but we can make them evaporate by coupling the CFT to an external system that can drain off the black hole energy.  The rate of this process is controllable by our choice of boundary conditions/couplings to the bath, and so we can always make the evaporation adiabatic\footnote{In this paper we will always assume we're dealing with a black hole that's in a pure state with characteristics typical of the micro-canonical ensemble.  Once the black hole gets small enough, this implies\cite{gary} that there is a phase transition to a higher dimensional Minkowski black hole, which evaporates into radiation rapidly before that radiation makes contact with our external bath. The questions we will study in this paper never enter into that regime. Our black holes are large compared to the AdS radius, both before and after the Page time.}.  

The island formula is derived by applying the replica wormhole formalism to the Euclidean gravitational path integral, using effective field theory to calculate one loop corrections to saddle point contributions.  In fact, at the level at which the calculations are done, one just assigns a field theoretic entropy to the radiation emitted into the bath. The crucial point is that the integral over metrics has two saddle points and they switch dominance at exactly the point at which the entropy in the bath is equal to the residual entropy on the horizon.   This is the point, beyond which the quantum information in the black hole can be read off from the radiation with higher and higher probability, according to the rules of quantum mechanics.  From this point on the entropy in the replica wormhole calculation is carried by the island and it is equal to the entropy of the radiation, and goes to zero as the evaporation continues.

\section{Where in the Island is the Entropy?}

Consider any coordinate system regular at the horizon on a time slice just before (say a few Planck times) the Page time, and consider points $\vec{x}$ on this slice, which will evolve into points in the island.  Now consider some geodesic $\vec{x} (\tau)$, starting at $\vec{x}$, and its complete causal diamond until it hits the singularity.  Recall that for large AdS black holes, the infall proper time to the singularity is never more than $R_{AdS} \ll R_S$, where $R_S$ is the Schwarzschild radius of the black hole.  Thus, for a typical point $\vec{x}$ and a typical geodesic, the causal diamond will have a maximal area on its boundary that is much much smaller than the entropy of the island at the Page time, which is of order $R_S^{d-2}$ in $AdS_d$.  The only way to avoid this is to start very close to the horizon with an outward pointing, almost null velocity.  Fortunately, there are points on the island that are quite close to the horizon.  In fact, the largest value of radial coordinate for the island is $r \sim r_H - K L_P$, where $K$ is the number of quantum fields that are used in computing the generalized entropy\cite{entanglementwedge} .    Bousso and Penington\cite{BP} have also argued that for higher dimensional black holes in Minkowski space, when angular momentum modes are taken into account, the island extends far outside the horizon.  These modes don't get out in the Hawking radiation, but a subset of them can be extracted from the black hole by physically plausible machinery.    

If we imagine detectors following these different time-like geodesics and believe in the covariant entropy bound as a bound on the maximal amount of quantum information that a detector inside a causal diamond could possibly access, these considerations tell us that all the quantum information in the island is contained in the part of it that skirts close to the adiabatic horizon of the evaporating black hole.  If one believes, as the present author does, that gravitational path integrals are only devices for capturing the coarse grained information of the fluctuating hydrodynamics of quantum gravity\cite{flucthydro} then one might be suspicious of attributing any information content larger than about $(R_{AdS}/L_P)^{d-2}$ to the trajectories outside the {\it stretched horizon}.  

We in fact believe that the entropy is concentrated in a much smaller region, of order the Planck scale or stretched horizon scale and we'll give an argument for this below.  This is not consistent with the bulk field theory description of near horizon thermodynamics on scales smaller than the AdS radius.  Although there is plenty of evidence that the bulk field theory provides a good description of few particle scattering amplitudes in the ``arena" causal diamond of Polchinski and Susskind\cite{polchsuss}, there is also ample evidence\cite{tHooftitaliansfirewall} that it does not provide a good description of high entropy states on such small diamonds.  

\section{Quantum Statistical Mechanics, Tensor Network Renormalization and Scale Radius Duality}

While many authors in the AdS/CFT literature view Error Correcting Codes/Tensor Networks\cite{swingle}\cite{adh}\cite{happy} as ``toy models" of AdS/CFT, many condensed matter physicists consider them to be a regularization/approximation scheme for CFTs.  The most sophisticated version of this point of view is the Tensor Network Renormalization Group (TNRG) of Evenbly and Vidal\cite{ev}.  One arranges a sequence of lattice field theories in shells, with one lattice point at the center of each close packed sphere on a hyperbolic space of dimension $d - 1$.  The successive shells are related by entanglement maps, which are supposed to make them approximations at longer and longer length scales to a lattice model which sits at a critical point and whose long wavelength physics is thus conformally invariant.  The longer wavelength models have fewer lattice points, so sit at smaller radius, implementing Maldacena's scale/radius duality in a very explicit way.  Calculations on simple models show that the small lattice models have finite matrix Hamiltonians that approximate the low lying dimension spectrum of the corresponding CFTs.  

Although carrying out the TNRG program for a holographic CFT is probably impossible in practice, there is no reason that it should not work in principle.  One can even imagine that one could mimic the success of\cite{happy} and make the tensor network invariant under an infinite discrete subgroup of translations of the hyperbolic space.  We want to ask what the wavefunction of a pure state that was picked to have the generic properties of the microcanonical ensemble represented by a black hole of radius $R_S \gg R_{AdS} \gg (L_P, L_s) $.  In particular, its probability distribution in the energy basis should be peaked at the black hole mass $M$ and have a width of order 
\begin{equation} \frac{\delta E}{M} = \sqrt{\frac{d - 2}{S}} , \end{equation} where $S$ is the black hole entropy, which scales as $(R_S / L_P)^{d-2}$ .    We can combine this with our knowledge that the density of states in CFT increases exponentially with a power of the energy, to conclude that the state can only be described in a small fraction of the lattice models in the tensor network.  Models with too small a radius do not contain the bulk of the states on which the probability weight of our pure state is concentrated.  Models with too large a radius do contain all of those states, but they also contain a huge number of states that are completely orthogonal to our chosen wave function.  Thus, from the point of view of the tensor network, the wavefunction of the black hole state is concentrated in a small radial interval, around the Schwarzschild radius.  Shells of the tensor network at larger radius just distribute the quantum information about the critical radius shell in an obscure way among a huge number of other degrees of freedom.  This is of course the whole point of an error correcting code.  Local deformations of the state of the larger shell will not affect the information about the state of the protected code subspace. 

What happens to the wave function as the black hole shrinks and the radiation bath becomes entangled with it? Obviously the position of the maximum probability moves.  Coarse grained logic would suggest that the energy distribution broadens because the black hole entropy has gone down. But we must remember that our setup is that we started from a fixed black hole wave function and extracted all of its quantum information into the radiation in the external system.  In particular, the radiation contains a record of the original narrow range of the energy distribution.  Since the energy extraction process is adiabatic, with a controllable rate, we can ensure that it has not broadened the energy envelope of the probability distribution.  The measured evaporated black hole is no longer a typical representative of its canonical distribution since it has a narrower than expected distribution of energies.  

This conclusion is consistent with the island formula conclusion, that after the Page time the island becomes entangled with the radiation, only if we admit that all of the degrees of freedom in the island are located in a small radial shell, mirroring the argument based on the covariant entropy bound.  One should be careful about the meaning of the word small in this discussion.  One has to translate the width in energy into length units using scale radius duality, which is less than precisely defined.  

One of the interesting side issues raised by this discussion is our claim that the evaporated black hole might have a narrower energy distribution than we'd expect from a typical pure state in the lower mass ensemble.  This is based on a very idealized picture of the Hawking radiation process and the process of ``measuring" the state of the outgoing radiation.  We are first of all assuming precise unitary evolution of the original pure state into a pure state of radiation.  This of course only happens if the black hole is completely isolated from any environment during the entire Page time.  Even more unlikely is our ability to measure a complicated quantum state.  This ALWAYS requires doing repeated experiments on identically prepared copies of the state in which semi-classical q-bits are entangled with eigenstates of different non-commuting q-bit operators.  Obviously this has no relation to any real experiment that one could ever do on a black hole.  A real black hole, starting from a typical pure state of the ensemble that is represented by its classical metric, and subjected to very minor environmental influences over the course of the Page time, will evolve into a mixed state dominated by typical pure states of the lower mass black hole ensemble.  

Even if we could keep our black hole state both pure and ``virginal", the chance that we could ever really do a complete quantum measurement of it is vanishingly small.  The best we could do with a single measurement on the radiation is to correlate a complete set of commuting Pauli operators on the radiation Hilbert space with some set of semi-classical q-bits in our measuring apparatus.  In doing so, we lose knowledge of relative quantum phase information between the eigenstates of the operators we've measured.  To complete the measurement we'd have to find another black hole, in exactly the same quantum state, let it evaporate, and measure a complementary set of operators.  The probability of finding a black hole in exactly the same state is something like $e^{-S_{BH}}$ and even if we had one, there's no way to know it's in exactly the same state.
Thus, while that questions we ask about the black hole information paradox can be made mathematically precise and have mathematical answers, those answers have nothing to do with real physical experiments on black holes.

\section{Islands Far Outside the Horizon?}

In an interesting paper, Bousso and Penington\cite{BP} argued that the ``island" region from which black hole information is retrieved by the Page time extends outside the horizon to a distance of order $r_S^{2/d}L_P^{(d-2)/d}$, where $r_S$ is the Schwarzschild radius of a black hole in asymptotically flat space of dimension $d \geq 4$.  The argument of\cite{BP} used the fact that although modes of higher angular momentum do not escape (with significant probability) in the Hawking radiation, the information in them can be ``mined" by dropping in finite tension strings from infinity.   In\cite{tbpdback} we argued that a similar conclusion could be reached by considering the semi-classical back reaction of the quantum state of effective field theory in the vicinity of the black hole horizon.  That is, if the state after the Page time is significantly different than the Hartle-Hawking state, then unless that difference is spread out over a region of order $r_S^{2/d}L_P^{(d-2)/d}$ it will make a large deviation in the mass of the black hole.  Similar conclusions about the kinds of states that would not back-react on the black hole mass were reached long ago\cite{thooftitalians}\footnote{In the author's opinion, these early papers should have been taken as strong evidence that the idea of accounting for black hole entropy in terms of field theory states with a ``brick wall" cut-off was incorrect.}.  

If a similar conclusion were true for large black holes in AdS space it would contradict our estimates of where the information in islands and black hole horizons is concentrated  for these black holes.  The Bousso-Penington argument does not apply directly to large AdS black holes, which are stable in the absence of couplings of the CFT to external systems.  The fate of angular momentum modes in this ``induced Hawking radiation" then depends on the nature of the coupling to the external system.  If the external system can extract angular momentum from the CFT, then the Hawking radiation will not be ``s-wave".  The back reaction argument of\cite{tbpdback} does not suffer from this problem.  It requires only an analysis of the behavior of bulk QFT states near the horizon of the large black hole, and the semi-classical back-reaction that they induce.  From the point of view of the dual CFT, large black holes are the extreme high energy spectrum, which is dominated by UV modes of the CFT.  

The back reaction calculation gives a very different answer for large AdS black holes.  Intuitively one can understand this in the tensor network construction, where each node of the network represents a causal diamond whose size $R_{node}$ in Planck units scales to infinity with $N$ as a power slightly less than $R_{AdS}/L_P$.  Then the arguments of\cite{tbpdback} indicate that the post Page time field theory state, if it differs from the Hartle-Hawking state, must do so over a region of size $R_{node}^{2/d} L_P^{(d-2)/d}$ .  The horizon of the black hole is covered by $(r_s /R_{node})^{d-2}$ of these nodes, so that is the size of the ``island outside the horizon".  Notice that the distance outside the horizon is independent of $r_s$.  This is consistent with our statistical argument from field theory and tensor networks above, because the natural distance scale in a tensor network is $R_{node}$.  

A more precise version of this argument is the following.  The blackening factor of an AdS-Schwarzschild black hole is 
\begin{equation} f(r) = 1 - \frac{\mu}{r^{d-3}} + \frac{r^2}{R_{AdS}^2} . \end{equation}   The horizon is at
\begin{equation} \mu = r_+^{d-3} (1 + \frac{r_+^2}{R_{AdS}^2}) . \end{equation}
For $r_+ \gg R_{AdS}$ the Hawking temperature is approximately $T \approx \frac{(d - 1) r_+}{4\pi R_{AdS}^2} $.  Near the horizon, the proper distance from the horizon is
\begin{equation} \rho = 2 \sqrt{( r - r_+)/f^{\prime} (r_+)} , \end{equation} so the local Tolman temperature
\begin{equation} T/\sqrt{f} = 1/2\pi\rho , \end{equation} independent of the black hole radius.  The coordinate depth of a shell of depth $\rho$ is $\delta r \sim \pi T \rho^2$.  

When expressed in these variables, the near horizon mode count of\cite{tbpdback} goes through unchanged.  If we consider a mode $e^{- i \omega t} Y_l \phi (r) $ for a massless scalar in leading order WKB approximation, the phase accumulates at the radial momentum
\begin{equation} k_{\rho} = \sqrt{\omega^2 /f - (l/r_+)^2}. \end{equation} If we put an 't Hooft brick wall at $\rho_h$ and sum phase space cells over the horizon, defining $\vec{k}_{\perp} = \vec{\Omega} k_{\perp}$ with $0 \leq k_{\perp} < \omega / \sqrt{f}$ , we find the number of modes
\begin{equation} N(\omega ) = \frac{A}{\pi (2\pi)^{d-2}} \int d^{d-2} k_{\perp} \int_{\rho_h} d\rho \sqrt{\frac{\omega^2}{f} - k_{\perp}^2} . \end{equation}   $A$ is the horizon area.  Up to order one numbers this gives
\begin{equation} N(\omega) = (\frac{\omega}{T})^{d-1} (\frac{r_+}{\rho_h})^{d-2} . \end{equation}
The density of states $dN/d\omega$ multiplied by the Bose-Einstein distribution gives us an energy carried by this state
\begin{equation} = \int d\omega \frac{dN}{d\omega} \frac{\omega}{e^{\omega/T} - 1} \sim T(\frac{r_+}{\rho_h})^{d-2} \sim \frac{r_+^{d-1}}{R_{AdS}^2 \rho_h^{d-2}} . \end{equation}
The corresponding entropy of this region is $S \sim \frac{A}{\rho_h^{d-2}}$ motivating the usual choice $\rho_h \sim L_P$ in the brick wall model's attempt to explain black hole entropy. 

Following\cite{tbpdback} we consider deviations from the Hartle Hawking density matrix in which an order one fraction of the modes are depleted or overexcited relative to the thermal ensemble.  Since the integral above is dominated by the lower endpoint $\rho_h$ the associated change in Schwarzschild frame energy is $\Delta U \sim \pm U_0 (\rho_h )$.  This causes the geometry outside the shell where we've altered the state to shift its redshift factor by
\begin{equation} \delta f \sim \frac{16 \pi L_P^{d-2}\Delta U}{(d - 2) \Omega_{d-2} r_+^{d-3}} . \end{equation} When $\delta f \sim f$ in the shell $f(\rho_h) = (2\pi T \rho_h)^2$, where the dominant modes sit, the local Tolman temperature of each of these modes has changed by a fraction of order $1$.   In terms of the actual calculation we have done, the radial turning points of the WKB wave functions move by more than the depth of the shell, so we were no longer including the correct modes in our calculation of $U_0$ and $\Delta U$.  The correction to these calculations is as big as the quantities themselves and the semi-classical expansion has completely broken down.  This is independent of whether we've depleted or overexcited the modes relative to the thermal HH state.  

The only way we can avoid this is to make $\rho_h$ large.  The condition that $\delta f < f$ around $\rho_h$ is
\begin{equation} \frac{L_P^{d-2} r_+^2}{R_{AdS}^2 \rho_h^{d-2}} < T^2 \rho_h^2 \sim (\frac{r_+}{R_{AdS}^2)^2} . \end{equation}  $r_+$ drops out completely and we find 
\begin{equation} \rho_h \sim (L_P^{d-2} R_{AdS}^2)^{1/d} = R_{AdS} (L_P/R_{AdS})^{\frac{d-2}{d}}) . \end{equation}  Interestingly, while this is smaller than the AdS radius, it is larger than the Planck scale by a power proportional to the ``number of fields" in the boundary CFT, which is the real measure of black hole entropy.  The total entropy of these modes scales like the area of the horizon, but in units of $\rho_h$ rather than $L_P$, suggesting that most of black hole entropy cannot be understood as coming from bulk QFT modes.  

The entropy count above is consistent with the tensor network prescription proposed in\cite{tbsadoublescaled}.  In that picture, each lattice point of the tensor network is identified with a causal diamond of size slightly smaller than $R_{AdS}$, whose entropy is described by a $1 + 1$ dimensional CFT that lives on the diamond's stretched horizon,following the prescription of Carlip and Solodukhin. 
The central charge is proportional to the area of the diamond in Planck units, which is roughly the same as the ``number of fields on the AdS boundary".  We argued that the bulk field theory could justify this picture but only gave an order $1$ central charge, corresponding to the number of independent bulk fields\footnote{Kaluza-Klein modes from the compact dimensions are not counted.  These just renormalize the higher dimensional Newton constant to the one used in $AdS_d$.}.   

\section{Non-isometric Encodings and Hilbert Bundles}

It's long been known\cite{cknetal} that QFT vastly overcounts the number of allowed quantum states in a model that admits black holes.  In fact, in flat space, it can only account for a fraction $S_{BH}^{(d-1)/d}$ of the black hole entropy allowed in a region the size of a black hole.  A clever way to account for the utility of QFT in approximating quantum gravity is to use the idea of {\it non-isometric encoding}\cite{noniso}:  the QG Hilbert space is given multiple descriptions, which are redundant, and together correspond to a field theoretic description of the physics.   A particular form of this very general framework is the Hilbert Bundle formulation of quantum gravity\cite{hilbertbundles}.  This is based on the view that classical solutions of Einstein's equations give a hydrodynamic description of certain states of underlying QG models.  They provide a division of the system into subsystems, and a prescription for some of the properties of the density matrices of those subsystems.  The subsystems are causal diamonds and so the natural non-isometric encoding is an encoding of a bundle of quantum systems over the space of time-like geodesics into a single QG Hilbert space.  The key properties of the encoding are strictly causal propagation along each geodesic, and a connection that relates the quantum information in diamonds along different geodesics by insisting that they give rise to the same entanglement spectrum for the density matrix on (the maximal diamond in) their overlap.  

The implication of our previous discussions for geodesics in the island region of a black hole interior are particularly interesting.  For all but the nearly null trajectories near the black hole horizon, the total size of their Hilbert space is bounded by something of order $R_{AdS}^{d-2}$.  Furthermore, if one looks at diamonds whose past tip is closer and closer to the singularity, the part of the Hilbert space accessible to future detection shrinks.  Thus, most of the quantum information on those diamonds must be detectable only by detectors on other geodesics, after a time of order $R_{AdS}$.  The connection on the Hilbert bundle lets us transfer this information up the Einstein-Rosen bridge (see below).  Note that this is not physical propagation of information faster than the speed of light.  It is more like a quantum mechanical version of the Newtonian and Coulomb force laws that allow one to ``keep track of" distant particles. 

Another way to think about it is by thinking about a geodesic orbiting the black hole.  The infalling geodesic that passes through the island, began life outside the black hole.  If the orbiting and infalling detectors have clocks synchronized at some time before the infalling detector passed through the horizon, then the proper time it takes for the infalling detector to hit the singularity is related to proper time on the orbiting detector by a modest redshift factor.  The orbiting detector never sees the infalling detector pass through the horizon.  Rather it sees phenomena on the horizon that can be interpreted as hydrodynamic traces of thermalization of the infalling detector subsystem with the larger black hole.  The purpose of the Quantum Principle of Relativity (QPR), the connection on our Hilbert bundle, is to make the quantum descriptions of all events compatible, independently of the fiber of the bundle on which we choose to do our quantum evolution.  Thus the consistency conditions between an infinite sequence of infalling diamonds are used to make the QM along the infalling trajectory as it hits the singularity compatible with the thermalization that is evident from the orbiting trajectory. 

The upshot of these results is that, at least for large AdS black holes where the island picture has been most reliably established, the quantum information in the island lies on geodesics the remain a distance from the horizon that is less than of order $ R_{AdS} (L_P/R_{AdS})^{\frac{d-2}{d}}) $, independently of the black hole radius $r_s \gg R_{AdS}$.   

\section{An Information ``Bucket Brigade" on the Einstein-Rosen Bridge}

Traditional discussions of information propagation in black hole interiors focus on information propagation by transmission of elementary particles.  One says that the Einstein-Rosen bridge expands faster than the speed of light, so that information encoded in local detectors that fell into the black hole can never be recovered classically.  This is really what leads to the paradox of information retrieval from the Hawking radiation.  The holographic perspective on quantum information throws this in a whole new light.  In this view, information is encoded on the maximal area surface on causal diamonds, and only of order $(A/G_N)^{(d-1)/d}$ of it can be realized in terms of bulk QFT excitations.  

Now consider the causal diamond of a detector thrown into a flat space black hole, as a function of the proper time of its past tip, starting at the time when its geodesic crossed the horizon. The maximal area on the diamond boundary gets smaller and smaller, which is of course the Hawking-Penrose definition of the singularity.  However, there's something interesting that occurs even if we look at the full diamond from horizon crossing to encounter with the singularity.  The past boundary of the diamond of course hugs the horizon of the black hole, but a pronounced ``waist" develops at a specific diamond area.
This maximum occurs at the midpoint of the trajectory towards the singularity and the waist has an area
\begin{equation} A_{waist} = (\frac{1}{2})^{\frac{d-2}{d-3}} . \end{equation}
For all finite $d$ this is less than $1/2$.   As a consequence, if we consider another detector dropped into the black hole slightly later than the first, the maximal area diamond in the region of overlap between the two detector diamonds will be greater than the area of the first diamond.

Recalling Page's theorem, and postulating\cite{lshpss} that information scrambles rapidly on diamond boundaries, we see that all the information on the first diamond is transferable to the second.  In this way we can construct a causal {\it holographic bucket brigade} which transmits information from any infalling detector diamond to the horizon.  Now all we must assume is that near the singularity, local degrees of freedom thermalize with the rest of the holographic variables on their tiny causal diamond, and we have a causal chain of information transmission from the depths of the ER bridge to the horizon.

This discussion, where calculations were done for flat space black holes, is also valid in most of the small causal diamonds from horizon crossing to encounter with the singularity, in large AdS black holes.  It implies that the information in the island is really concentrated in the interior within a Planck distance from the horizon.  To see this, note that the geometry inside causal diamonds approaching the singularity is completely determined by the ratios between proper time and maximal area for diamonds whose proper time before the singularity is $\tau$.  Along radially infalling geodesics the two functions $A(\tau)$ are obviously the same right near the singularity at $\tau = 0$.  In AdS space, the expansion away from $\tau = 0$ is
\begin{equation} A(\tau) = 4\pi (9M/2)^{2/3} \tau^{4/3} [1 - \frac{2 (3\tau/2)^{2/3}}{3 (2M)^{2/3}} + \frac{2 (2M)^{2/3} (3\tau/2)^{4/3}}{3 L^2} + \ldots ] . \end{equation} In flat space the last term and the ellipsis are absent.  The condition that the metric inside the diamond match that inside a flat space black hole of Schwarzschild radius $\sim L$ is thus that $\tau < L$.  We've restricted this calculation to $d = 4$ and measured everything in Planck units.  The calculation can be generalized to geodesics with angular momentum, and as long as $J_z < L/L_P$ the conclusion is the same.  Similarly, if we consider two geodesics that enter the same region of area $L^{d-2}$ on the horizon with a proper time delay that's less than $L$, the difference between the flat space and AdS calculations is again negligible.  We conclude that the holographic buket brigade transports information to within a distance $\ll L $from the horizon for large AdS black holes.  The island computation is really only sampling information about the completelu
equilibrated state of the black hole at the Page time.  EFT seems to suggest that some of that information is located at causally inaccessible regions along the ER bridge, but this is because EFT fails to capture the true holographic dynamics of quantum gravity.  

\section{Conclusions}

In the first part of this paper, we used arguments from bulk EFT and the tensor network picture of boundary EFT to conclude that the information in the ``island" region of a large AdS black hole was concentrated in a region whose size was of order at most the AdS radius around the black hole horizon.   In the last section we used somewhat more speculative arguments to show that in fact, for all black holes, information is rapidly transmitted to a Planck scale stretched horizon.  These arguments give support to the idea that the encounter with the singularity in a black hole is a sign of the thermalization of local degrees of freedom with the more numerous holographic q-bits of a model in which most quantum information lives on the boundaries of causal diamonds, and is inaccessible to local detectors.  In Minkowski space string theory these holographic q-bits become the ``soft supergraviton states", whose presence can be ignored above $4$ dimensions and at finite energy.  In AdS space, these states are sent off to infinite energy by the process of renormalization.  Finite energy remnants of them exist as the dominant contributions to the entropy of large AdS black holes, which cannot be understood (microscopically) in terms of the bulk AdS field theory of BDHM/HKLL\cite{bdhmhkll} and\cite{LL}.

\section*{Acknowledgments}

The author thanks A. Almheiri, Gemini,Claude, and Aidan Herderschee for useful discussions of the material in this paper.  Some of the calculations of the detailed properties of large AdS black holes are due to Patrick Draper, who has generously allowed me to include them here.  This paper was written while the author was on sabbatical at the Center for Theoretical Physics at MIT - a Leinweber Institute. I'd like to thank the institute for its hospitality.  

\end{document}